\documentclass{solarphysics}
\usepackage[hyperref,optionalrh,solaromanenum]{spr-sola-addons} % For Solar Physics 
\usepackage{epsfig}                     % For eps figures, old commands
\usepackage{graphicx}                    % For eps figures, newer & more powerfull
\usepackage{amssymb}
\usepackage{hyperref}
\usepackage{amssymb}                    % useful mathematical symbols
\usepackage{color}                       % For color text: \color command
\usepackage{url}
\usepackage{breakurl}                         % For breaking URLs easily trough lines
\usepackage{natbib}

\newcommand{\etal}{{\it et al.}}

\newcommand{\degr}{\hbox{$^{\circ}$}}

\begin{document}

\begin{article}

\begin{opening}

\title{Testing and Improving a Set of Morphological Predictors of Flaring Activity}

%%%%%%%%%%%%%%%%%%%%%%%%%%%%%%%%%%%%%%%%%%%%%%%%%%%
%% Authors Names
%
% \author[addressref={},corref,email={}]{\inits{}\fnm{}\lnm{}\orcid{}}

\author[addressref=aff1,corref,email={jkonto@noa.gr}]{\inits{I.}\fnm{Ioannis}~\lnm{Kontogiannis}~\orcid{0000-0002-3694-4527}}
\author[addressref=aff1,email={manolis.georgoulis@academyofathens.gr}]{\inits{M.K.}\fnm{Manolis K.}~\lnm{Georgoulis}~\orcid{0000-0001-6913-1330}}
\author[addressref=aff2,email={sunpark@tcd.ie}]{\inits{S.-H.}\fnm{Sung-Hong}~\lnm{Park}~\orcid{0000-0001-9149-6547}}
\author[addressref={aff3,aff4},email={guerraaj@tcd.ie}]{\inits{J.A.}\fnm{Jordan A.}~\lnm{Guerra}}%\sep

\address[id=aff1]{Research Center for Astronomy and Applied Mathematics (RCAAM) Academy of Athens, 4 Soranou Efesiou Street, Athens, GR-11527, Greece}
\address[id=aff2]{Institute for Space-Earth Environmental Research, Nagoya University, Nagoya, Japan}
\address[id=aff3]{School of Physics, Trinity College Dublin, Ireland}
\address[id=aff4]{Physics Department, Villanova University, Villanova, PA, USA}

%\author{Ioannis~\surname{Kontogiannis}$^{1}$\sep
%        Manolis~K.~\surname{Georgoulis}$^{1}$\sep
%        Sung-Hong~\surname{Park}$^{2}$\sep
%        Jordan~A.~\surname{Guerra}$^{3}$\footnote{Now at Physics Department, Villanova University, Villanova, PA, USA}\sep          }
%%%%%%%%%%%%%%%%%%%%%%%%%%%%%%%%%%%%%%%%%%%%%%%%%%%
%% Runningheads
%
\runningauthor{Kontogiannis \etal}
\runningtitle{Morphological predictors of flaring activity}

%%%%%%%%%%%%%%%%%%%%%%%%%%%%%%%%%%%%%%%%%%%%%%%%%%%
%% Affilations
%% id shold be the same with \author addressref value.
%\address[id={}]{}
%  \institute{$^{1}$ Research Center for Astronomy and Applied Mathematics (RCAAM) Academy of Athens, 4 Soranou Efesiou Street, Athens, GR-11527, Greece
%                     email: \url{jkonto@noa.gr} \\
%             $^{2}$  Institute for Space-Earth Environmental Research, Nagoya University, Nagoya, Japan \\
%             $^{3}$  School of Physics, Trinity College Dublin, Dublin 2, Ireland                    \\}
%%%%%%%%%%%%%%%%%%%%%%%%%%%%%%%%%%%%%%%%%%%%%%%%%%%
%%% Abstract
\begin{abstract}
Efficient prediction of solar flares relies on parameters that quantify the eruptive capability of solar active regions. Several such quantitative predictors have been proposed in the literature, inferred mostly from photospheric magnetograms and/or white light observations. Two of them are the Ising energy and the sum of the total horizontal magnetic field gradient. The former has been developed from line-of-sight magnetograms while the latter utilizes sunspot detections and characteristics, based on continuum images. Aiming to include these parameters in an automated prediction scheme, we test their applicability on regular photospheric magnetic field observations provided by the \textit{Helioseismic Magnetic Imager} (HMI) instrument onboard the \textit{Solar Dynamics Observatory} (SDO). We test their efficiency as predictors of flaring activity on a representative sample of active regions and investigate possible modifications of these quantities. Ising energy appears as an efficient predictor, with efficiency improved if modified to describe interacting magnetic partitions or sunspot umbrae. The sum of the horizontal magnetic field gradient appears as slightly more promising than the three variations of the Ising energy we implement in this work. The new predictors are also compared with two very promising predictors, the effective connected magnetic field strength and the total unsigned non-neutralized current. Our analysis shows that the efficiency of morphological predictors depends on projection effects in a nontrivial way. All four new predictors are found useful for inclusion in an automated flare forecasting facility, such as the Flare Likelihood and Region Eruption Forecasting (FLARECAST), but their utility, among others, will be ultimately determined by the the validation effort underway in the framework of the FLARECAST project. 
\end{abstract}

%%%%%%%%%%%%%%%%%%%%%%%%%%%%%%%%%%%%%%%%%%%%%%%%%%%
%% Keywords
%
\keywords{Active Regions, Magnetic Fields; Flares, Forecasting; }

\end{opening}
%-------------------------------------------------

%%%%%%%%%%%%%%%%%%%%%%%%%%%%%%%%%%%%%%%%%%%%%%%%%%%
%% Sections
%
% \section{}%\label{s:?}
\section{Introduction}
\label{s:intro}

Flares are impulsive releases of energy in the solar atmosphere, manifested by increased emission throughout the electromagnetic spectrum \citep{fletcher11}. The importance of a solar flare is measured by its soft X-ray emission, as recorded by the \textit{Geostationary Operational Environmental Satellites} (GOES). Flare classes are denoted by letters A, B, C, M, and X, complemented by decimal sub-classes. Flares of C class and above are usually deemed more important in terms of space weather effects while M1.0 class and above are often characterized as major. Solar flares are often associated with acceleration of solar energetic particles (SEP) and/or coronal mass ejections (CME's), \textit{i.e.} blasts of coronal plasma into the interplanetary space \citep[see \textit{e.g.,}][and references therein]{tziotziou10} . 

Much of the impact of solar flares is immediate. The UV and X-ray radiation of solar flares affects immediately the electron density of the ionosphere on the day-side \citep{hayes17}, causing radio disruptions, expansion of the atmosphere and increase drag on low-orbit satellites. Furthermore, increased radiation (both electromagnetic and particles) can be harmful to instruments and crew in space.	Thus, to mitigate the impact of solar flares and related phenomena, we can only rely on accurate prediction methods.  

It is widely known that the energy that powers solar flares is stored in the magnetic field of solar active regions. These are hosts to strong and complicated magnetic concentrations, manifested in white light as sunspots and in the UV as a multitude of million-degree coronal loops. The morphology of the magnetic field configuration, inferred by photospheric line-of-sight (LOS) or vector magnetograms often indicates the potential of an active region to flare. The presence of strong magnetic polarity inversion lines (MPIL), when opposite magnetic polarities are found in close proximity and exhibit strong shearing motions are indicative of high non-potentiality of the active region magnetic field. These configurations are prone to major flaring activity \citep[see \textit{e.g.,}][and references therein]{schrijver16}.

In this sense, a number of morphological parameters have been developed to describe the magnetic field configuration and/or quantify the presence of MPILs in active regions. A non-exhaustive review of the pertinent literature can be found in \citet{georgoulis12}. Morphological parameters may be of varying sophistication. The total unsigned magnetic flux of an active region is the simplest morphological parameter, which quantifies its size. More sophisticated efforts have focused specifically on the MPILs, \textit{e.g.} either by parametrizing the total magnetic flux within a distance from the MPIL \citep[R-parameter, see][]{schrijver07} or its magnetic-flux-weighted length \citep[\textit{e.g.} ][]{falconer08}. Another approach is to take into account the connectivity between opposite-polarity magnetic elements, as done by \citet{georgoulis07} or consider the physical cause of shear and MPIL formation by measuring the amount of non-neutralized currents injected into the corona \citep{georgoulis12b,kontogiannis17}. In this study, we will test two more morphological parameters that rely on the connections between opposite-polarity magnetic elements and propose modifications to improve their efficiency.

The horizontal magnetic gradient has been introduced in flare prediction by \citet{korsos14}, \citet{korsos15}, and \citet{korsos16}. They utilized sunspot umbrae areas and positions from the Heliophysical Observatory, Debrecen database \citep{baranyi16} to develop parameters which describe the magnetic gradients within an active region. Two of them, $G_{M}$ and $WG_{M}$ require the manual selection of regions of interest within the active region (\textit{e.g.} near MPILs). The third one, $G_{S}$, \textit{i.e.} the sum of the horizontal magnetic field gradient, characterizes the entire active region and does not require the definition of regions-of-interest for its calculation. Therefore, it is suitable for implementation in an automated forecasting facility such as the one developed in the framework of the Flare Likelihood and Region Eruption Forecasting (FLARECAST) consortium.

The Ising energy is a complexity measure introduced by \citet{ahmed10}. They borrowed the term from the Ising model of ferromagnetism in solid state physics, which is used to describe the spin-spin interaction in a distribution of magnetic elements. The authors showed that active regions with larger values of Ising energy exhibited more and stronger flares. Here we adapt their algorithm, also implementing two variations, the Ising energy of the magnetic partitions and the Ising energy of the sunspot umbrae. The former refers to the interaction between opposite-polarity magnetic partitions while the latter is calculated for interacting opposite-polarity sunspot umbrae.

In this article we reproduce the methodologies described above and adjust them to the context of an automated flare forecasting service that relies on a regular flow of active region cutouts. Then, for the first time, we test the four parameters on a representative sample of observations and assess whether they are potentially efficient predictors of flaring activity. This is done by comparing their efficiency against that of the simplest possible morphological predictor, namely the total unsigned magnetic flux, as well as against two promising predictors, the effective connected magnetic field strength, $B_{eff}$ \citep{georgoulis07} and the total unsigned non-neutralized electric current $I_{NN,tot}$ \citep{kontogiannis17}.

\section{Data}
\label{data}

Observations from the \textit{Helioseismic Magnetic Imager} \citep[HMI:][]{hmischerrer,hmischou} onboard the \textit{Solar Dynamics Observatory} mission \citep[SDO:][]{sdo} were used. The HMI team have developed the Space Weather HMI Active Region Patches \citep[SHARP:][]{bobra14}, which comprise data suitable for space weather forecasting applications. SHARP partial-disk cutouts contain the magnetic field vector components of areas of interest, remapped and deprojected as if observed at the solar disk center along with a set of quantities used for predicting solar eruptions. In this study the near-real time (NRT), cylindrical equal area (CEA) SHARP data are used.

SHARPs are usually, but not always, accompanied by National Oceanic and Atmospheric Administration (NOAA) active region (AR) identifiers. The correspondence between the two is not one-to-one. A HARP (which refers to the cutout only, without the extra diagnostic [``space weather'', S] information) may contain one, several, or no NOAA active region numbers. In principle, regions that appear from the eastern limb are not recognized immediately and are assigned a number after traversing ten or twenty degrees in heliographic (HG) longitude. Also, active regions must reach a certain size to be visible by NOAA.

The Heliophysical Observatory, Debrecen database\footnote{http://fenyi.solarobs.csfk.mta.hu/en/databases/SOHO/} \citep{baranyi16} contains an archive of active region information that spans the operation years of the \textit{Michelson Doppler Imager} \citep[MDI:][]{mdi} and the HMI. It combines the corresponding magnetic field observations with ground-based continuum observations to deduce the number of sunspots, umbral/penumbral areas, positions, \textit{etc} for each active region on the Sun. From this database, the SDO/HMI Debrecen Sunspot Data (HMIDD) were produced by HMI observations that are relevant to our dataset. 

For the development and initial testing of the algorithms, we used two time series of NOAA ARs 11158 and 11923. These two active regions differ dramatically in terms of flaring productivity, the former being quite active and the latter producing no flares of C class and above. To further investigate and assess the potential of the morphological parameters as predictors of flaring activity we further processed a representative sample of SHARP cutouts of Solar Cycle 24. For each of the 336 random days between September 2012 and May 2016 we processed cutouts at a cadence of 6\,h, with no constrains on the solar disk location. This gives us also the opportunity to investigate the variation of the morphological predictors as a function of the heliographic longitude. Thus, the representative sample consisted of 9454 point-in-time observations of regions of interest.

The association of the regions observed by HMI with flaring activity within the following 24\,h was derived utilizing flare detections from the \textit{Geostationary Operational Environmental Satellite} (GOES\footnote{\url{http://www.ngdc.noaa.gov/stp/satellite/goes/}}) and more specifically from the NOAA Edited Solar Event Lists\footnote{\url{ftp://ftp.swpc.noaa.gov/pub/warehouse}}, which includes GOES flare start times, source regions, and locations. Out of the 9454 points of the sample, 1007, 138, and 15 are associated with C, M, and X class flares, respectively, over the preset window.

 \begin{figure}
 \centerline{\includegraphics[width=1\textwidth]{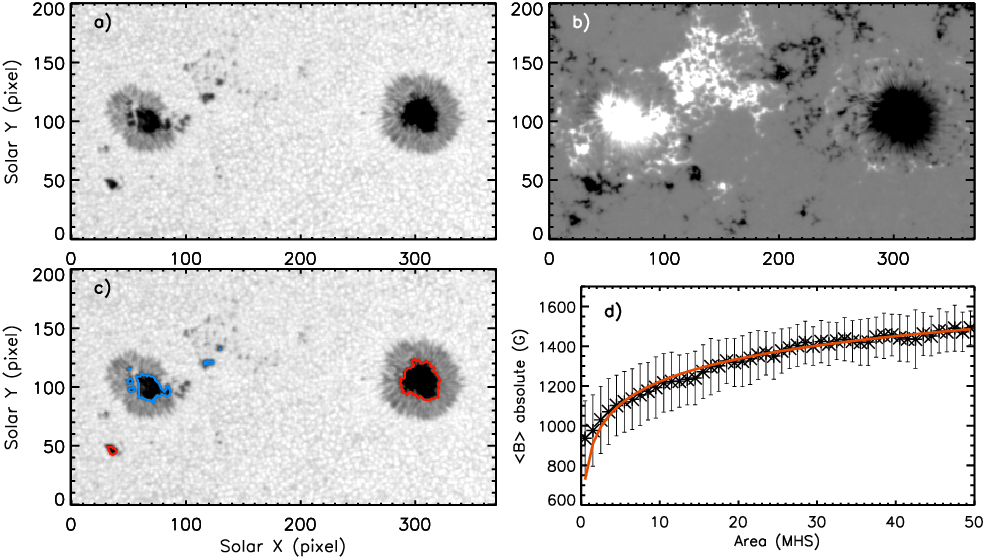}}
 \caption{a--b) Continuum image and co-temporal line-of-sight magnetogram of NOAA AR\,11611, taken in 12 November 2012, at 12:58\,UT. c) continuum image with overlaid contours of the positive (\textit{red}) and negative (\textit{blue}) umbral areas produced with the method of \citet{padinhatteeri16}. d) Absolute mean magnetic field as a function of umbral area (\textit{black stars}) and the overplotted fitted curve (\textit{red line}). The values are taken from the database of the Debrecen Observatory and are calculated for the HMI data between 2010 and 2014. Only active regions within $10\degr$ from the central meridian are considered.}
\label{fig:umbrae_det}
 \end{figure}

\section{Analysis}
\label{s:analysis}

\subsection{Sum of the Horizontal Magnetic Field Gradient}
The sum of the horizontal magnetic field gradient \citep{korsos16}, $G_{S}$, is given by the formula:

\begin{equation}
G_{S}=\sum_{i,j}\frac{B_{p,i}A_{p,i}-B_{n,i}A_{n,i}}{d_{i,j}},
\label{equation:gs}
\end{equation}

\noindent where $B_{p}$ ($B_{n}$) is the mean magnetic field of the positive (negative) umbrae and $A_{p}$ ($A_{n}$) is the corresponding area. The quantity $d_{i,j}$ is the distance between the flux-weighted centroids of the positive and negative magnetic polarity umbrae. The double summation runs for all possible positive- and negative-polarity umbrae. Continuum images are used to automatically extract umbral and penumbral areas and positions \citep{gyori98} and the LOS magnetograms are used to correspond each umbral area, $A$, to a mean magnetic field $<B>$. An empirical relationship between the umbral area and the mean magnetic field is constructed by observations of active regions within $10\degr$ from the solar disk center \citep{korsos14}. The curve $<B> = f(A)$ is then fitted to produce the fitting relation: 

\begin{equation}
<B>=f(A)=K_{1}ln(A)+K_{2}.
\label{equation:gs_calibr}
\end{equation}

\noindent To minimize the effect of geometrical foreshortening and projection on the measurement of the magnetic field, only the umbral areas, corrected for center-to-limb variation, are used to calculate $G_{S}$ in Equation~\ref{equation:gs}. 

To calculate $G_{S}$ we use the line-of-sight magnetograms and continuum images provided with each HARP. These are remapped to the CEA coordinate system. We also correct the magnetograms by dividing the LOS component by the cosine of the position angle. It should be noted that, since the fitting relation is applied only to observations at the disk center, using the radial component would not lead to different results. We measure the areas and positions of the umbrae within a HARP cutout, following the simple approach of \citet{padinhatteeri16}, which is based on threshold values of intensity and magnetic flux density for umbrae and penumbrae areas (see Figure~\ref{fig:umbrae_det}a--c).
To convert the areas to mean magnetic field strength, we use the HMIDD database of the Debrecen Observatory, which contains sunspot detections based on HMI continuum images and line-of-sight magnetograms. In Figure~\ref{fig:umbrae_det}d we plot $<B>=f(A)$ for active regions located within $10\degr$ from the solar disk center. After performing a least-squares fit to the data points, we find $K_{1}=165$\,G and $K_{2}=842.16$\,G. These values are then used to convert umbral areas to mean magnetic field.

 \begin{figure}
 \centerline{\includegraphics[width=1\textwidth]{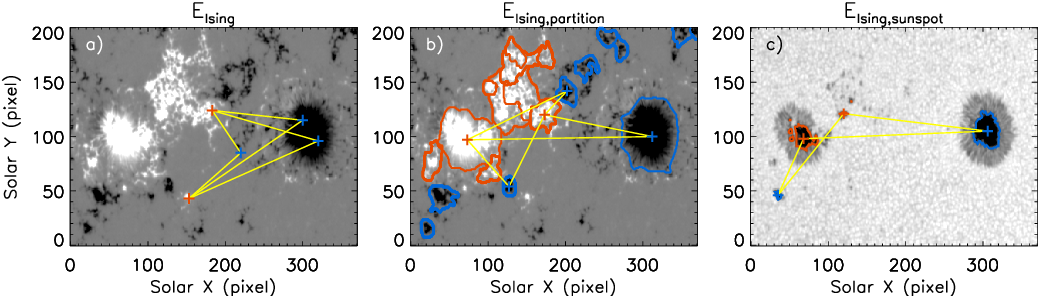}}
 \caption{From \textit{left} to \textit{right}: Example of Ising energy calculations for different types of magnetic elements distribution, \textit{i.e.} opposite polarity pixel, partitions of the magnetic field and sunspot umbrae.}
\label{fig:ising_calc}
 \end{figure}

\subsection{Ising Energy}
\label{s:ising_energy}
\noindent According to \citet{ahmed10} the Ising energy of a distribution of interacting magnetic elements (represented by image pixels) is calculated via the formula:

\begin{equation}
E_{Ising}=\sum_{i,j}\frac{S_{i}S_{j}}{d^{2}},
\label{equation:ising}
\end{equation}

\noindent where $S_{i}$ ($S_{j}$) equals to +1 (-1) for the positive (negative) magnetic polarity pixel of each pair and d is the distance between the two opposite polarity pixels (Figure~\ref{fig:ising_calc}a). Therefore, numerous opposite polarity pixels in small distances, \textit{e.g.} along MPILs, should contribute more to the Ising energy than highly separated ones. The LOS magnetograms are position-angle corrected and then smoothed by a 5-pixel running average along the x- and y-axis. The weak magnetic field strength pixels are removed from consideration by setting a threshold equal to 100\,G and, then, the Ising energy is calculated by means of Equation~\ref{equation:ising}.

Additionally to the original form of the Ising energy, we implement two variations, based on the original definition. The first one, the Ising energy of the partitioned magnetogram, $E_{Ising,part}$, refers to the ensemble of the pairs of opposite polarity partitions (Figure~\ref{fig:ising_calc}b), produced by the gradient based partitioning scheme of \citet{barnes05}. This method requires additional thresholds of the magnetic patch area and contained magnetic flux, which are set to 40 pixels and $5\cdot10^{19}$\,Mx respectively. The second one, the Ising energy of sunspot umbrae distribution $E_{Ising,spot}$, utilizes the sunspot detection method used for $G_{S}$ and is calculated for all the pairs of opposite polarity sunspot umbrae within an active region (Figure~\ref{fig:ising_calc}c). 

\begin{table}
\caption{Active regions 11923 and 11158. $t_{start}$ and $t_{end}$ are the starting and ending dates of each time series while long(start:end) are the corresponding heliographic longitudes in degrees. C, M, and X columns denote the number of the corresponding class flares within this interval and Mag. Type is the Mt.Wilson classification type of the NOAA AR.}
\setlength{\tabcolsep}{4.1pt}
\begin{tabular}{lccccccc}
\hline
NOAA AR&$t_{start}$\,(UT)&$t_{end}$\,(UT)&long(start:end)&C&M&X&Mag. Type\\
\hline
11158&2011-02-10 21:59&2011-02-15 22:59&     -42:25&  25&      4&       1&  $\beta$-$\beta\gamma$\\
11923&2013-12-12 03:35&2013-12-15 22:35&     -18:32&  0&       0&       0&  $\beta$\\
\hline
\end{tabular}
%\tablefoot{}
\label{Table:t1}
\end{table}

\subsection{$B_{eff}$, $I_{NN,tot}$, and $\Phi_{tot}$}

In the following we will compare the four new parameters with two other morphological parameters as well as the total unsigned magnetic flux, calculated for the same sample of observations. 

The effective connected magnetic field strength, $B_{eff}$, was introduced by \citet{georgoulis07} and it is a morphological parameter that emphasizes the presence of MPILs. It is already in use by automated flare forecasting services (Athens Effective Solar Flare Forecasting, A-EFFORT\footnote{\url{https://a-effort.academyofathens.gr}}, service by the European Space Agency's Space Situational Awareness Programme\footnote{\url{http://swe.ssa.esa.int/solar-weather}}), where it is calculated based on the angle-corrected LOS component of the magnetic field. We use the $B_{LOS}$ version in this article as well.

The total unsigned non-neutralized currents, $I_{NN,tot}$, is a new predictor of flaring activity, introduced by \citet{kontogiannis17} and it represents the net currents injected into the corona. It is exclusively linked with the presence of strong MPILs; active regions without MPILs contain, in principle, zero non-neutralized currents. $I_{NN,tot}$ is calculated using the three components of the magnetic field.

The total unsigned magnetic flux is the sum of the absolute flux contained within each pixel of the FOV. It is calculated from the radial component of the magnetic field, taking into account that the pixel area of the CEA NRT SHARP cutouts is 1.33$\cdot$10$^{5}$\,km$^{2}$. Only pixels with magnetic flux density higher than 100\,Mx$\cdot$cm$^{-2}$ were taken into account. 

\section{Results}
\label{s:results}

\subsection{Active Region Evolution and Parameter Time Series}
\label{s:ars}
As a preliminary examination, we calculate the evolution  of the four parameters for two active regions, AR\,11158 and AR\,11923. The former is a very active and a well studied one \citep[see, \textit{e.g.}][]{tziotziou13} while the latter produced no flares of C class and above. 

\begin{figure}
 \centerline{\includegraphics[width=1\textwidth]{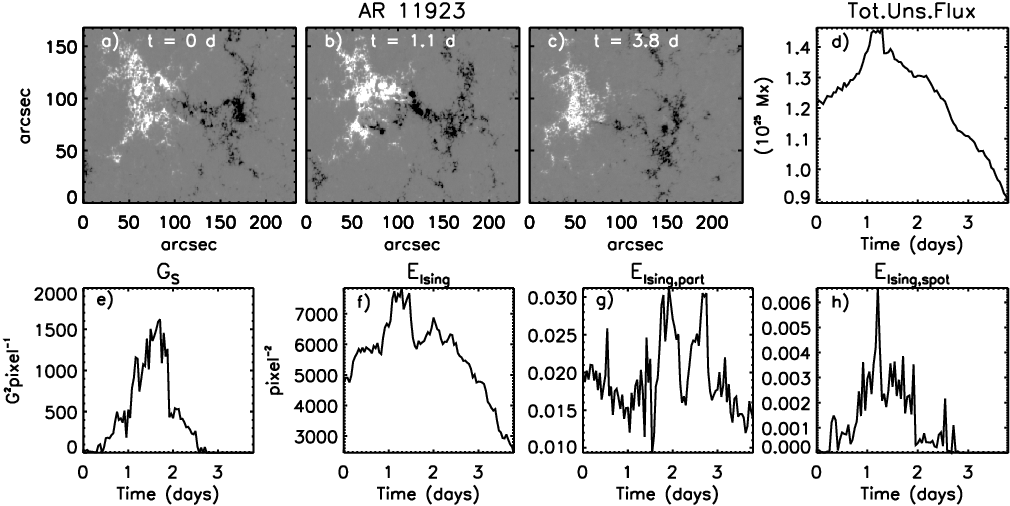}}
 \caption{Sample calculations of the four parameters for NOAA AR\,11923. a--c) Maps of the line-of-sight component of the magnetic field at the beginning, the time of the maximum total unsigned magnetic flux, and at the end of the observations. d) The temporal variation of the total unsigned magnetic flux during the observations. e--h) The corresponding time series of the sum of the horizontal magnetic gradient, the Ising energy, the Ising energy of the magnetic partitions, and the Ising energy of the sunspot umbrae.}
\label{fig:ar11923}
 \end{figure}

NOAA AR\,11923 (Figure~\ref{fig:ar11923}) was observed between 12 and 15 December 2013. In the continuum it consisted of several small sunspots and remained a $\beta$-type in the Mt.Wilson classification throughout its passage through the disk. Its total unsigned magnetic flux increased during 12 December, peaked at $\sim$1.4$\cdot 10^{22}$\,Mx on 13 December and then declined during the following days (Figure~\ref{fig:ar11923}d). Maps of the LOS magnetic field component, $B_{LOS}$, are shown in Figures~\ref{fig:ar11923}a--c for three instances, \textit{i.e.} at the beginning, the time of maximum flux, and the end of the observations.  The two polarities being well-separated, exhibited no strong MPILs, and no shearing motions. 

The corresponding time series reflect the evolution of the active region. $G_{S}$ and $E_{Ising,spot}$ (Figures~\ref{fig:ar11923}e and \ref{fig:ar11923}h respectively) are equal to zero at the beginning and at the end of the time series, \textit{i.e.} when there are no detectable umbrae in the continuum. The increase of flux and the appearance of opposite polarity umbrae separated by small distances produces the distinct peaks at the corresponding time series during the second day. The $E_{Ising}$ (Figure~\ref{fig:ar11923}f) closely resembles the total unsigned magnetic flux time series. As the magnetic flux increases numerous pairs of opposite polarity pixels appear contributing to the total value of $E_{Ising}$. The evolution of $E_{Ising,part}$ (Figure~\ref{fig:ar11923}g) is less conspicuous than the other three parameters. The active region consists of several relatively small partitions scattered across the FOV. Thus the presence of a few more partitions during the increase of the flux does not produce a very distinct variation in the time series.  

 \begin{figure}
 \centerline{\includegraphics[width=1\textwidth]{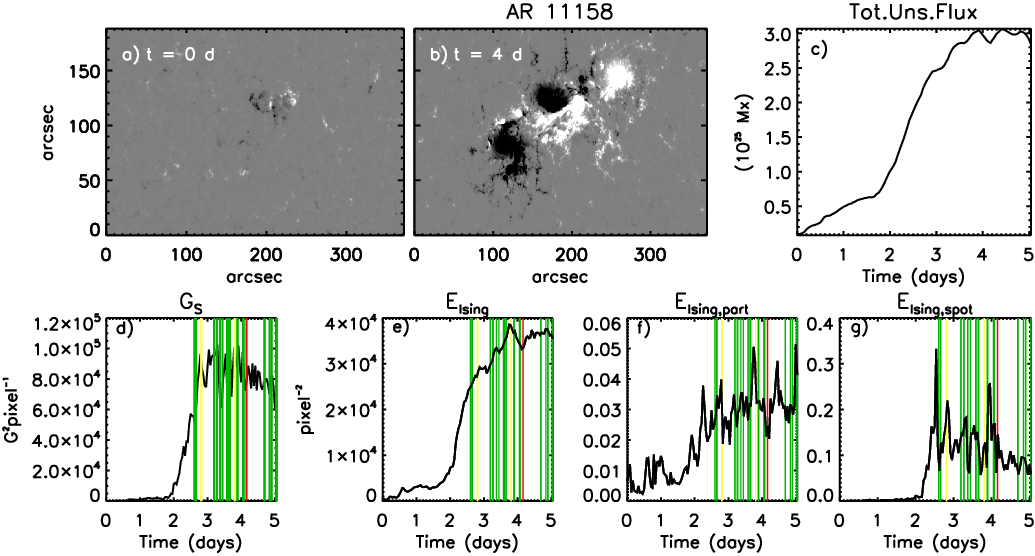}}
 \caption{Same as Figure~\ref{fig:ar11923} but for NOAA AR\,11158. In the panels of the bottom row, the \textit{vertical} lines denote the times of flare occurrences, \textit{green}, \textit{yellow}, and \textit{red}, standing for C, M, and X class flares, respectively.}
\label{fig:ar11158}
 \end{figure}

In Figure~\ref{fig:ar11158}, we present our calculations for the well-studied NOAA AR\,11158. This started as a small bipolar region (Figure~\ref{fig:ar11158}a) but then at the end of the second day, a dramatic emergence of flux turned this active region from $\beta$ to $\beta\gamma$. A quadrupolar configuration appeared (Figure~\ref{fig:ar11158}b), along with a strongly sheared MPIL. The active region produced several C and M class flares as well as the first X class flare of Solar Cycle 24. 

The evolution of the total unsigned magnetic flux and the morphology of the active region reflect upon the temporal variation of the four parameters. From the third day onwards, the increased values of the parameters are associated with the intense flaring activity of AR\,11158 that followed. Several peaks in the time series either precede or coincide with strong flares. In general, this flare-productive active region was more than twice as large, in terms of magnetic flux content, than AR\,11923. This also reflects on the values of the parameters, which are higher than the corresponding ones for AR\,11923. Specifically, $G_{S}$, $E_{Ising}$, and $E_{Ising,spot}$ are more than an order of magnitude higher than the corresponding values of AR\,11923. A statistical examination of the robustness of this result is presented in Section~\ref{s:flare_predi}.

\subsection{Parameter Distributions and Variation Across the Solar Disk}
\label{s:params_distr}

\begin{figure}
\centerline{\includegraphics[width=1\textwidth]{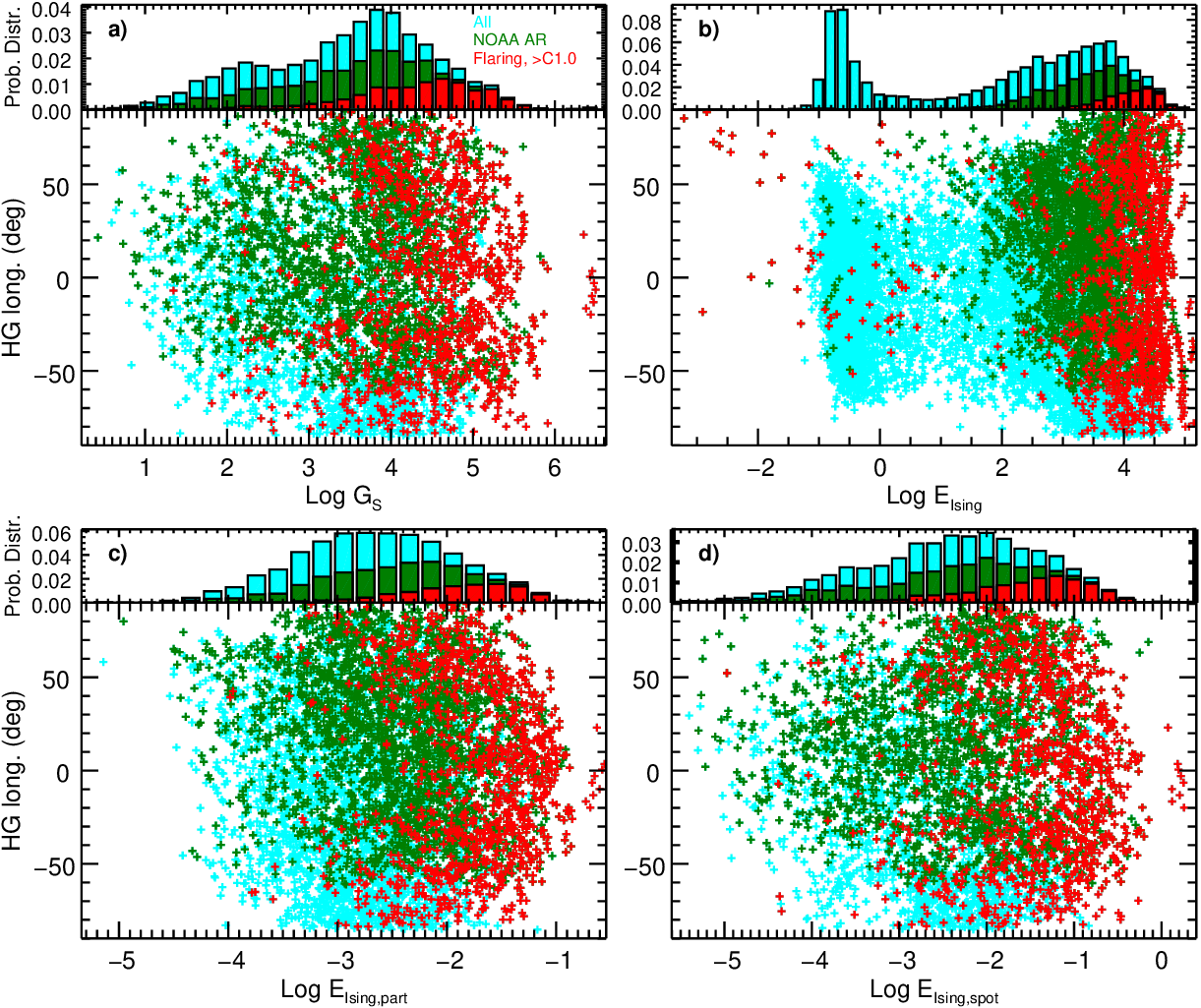}}
 \caption{Histograms of the values of the four parameters for the representative sample of HARPs. \textit{a)} $G_{S}$, \textit{b)} $E_{Ising}$, \textit{c)} $E_{Ising,part}$, and \textit{d)} $E_{Ising,spot}$. \textit{Blue} denotes the entire sample, \textit{green} refers to the HARPS with NOAA AR number and \textit{red} to HARP regions associated with at least a C class or higher flare within the following 24\,h. Each histogram is accompanied by a scatter plot that shows the distribution of the values as a function of the heliographic longitude.}
\label{fig:histos}
\end{figure}

In Figure~\ref{fig:histos}, we plot the histograms of the values of the four parameters for the 9454 SHARP point-in-time observations and their distribution on the solar disk with respect to the heliographic longitude of the HARPs. It should be noted that all four parameters produce zero values when either of the following cases apply: 1) there is only one polarity present in the FOV, 2) there are only weak-field pixels (lower than 100\,G), 3) there are no sizeable partitions, as per the thresholds applied (see Section~\ref{s:ising_energy}), or 4) there are no sizeable umbrae. A zero value in this case is meaningful since it represents areas where presumably no morphological features associated with flaring activity are present. In the following, we study the distribution of the non-zero values.

The distribution of the sum of the horizontal magnetic field gradient values, $G_{S}$, is overall asymmetric (Figure~\ref{fig:histos}a). It shows a dominant peak at $log G_{S}=3.8$ and a secondary one at $log G_{S}=2.2$. The corresponding distribution for HARPs with an assigned NOAA AR number is similar but lacking the secondary peak. As already mentioned, HARPs without an assigned NOAA number are regions either without sizeable sunspots and thus indiscernible in continuum observations or very close to the eastern limb. Regarding the latter, this is also obvious from the coresponding distribution with the heliographic longitude. The green crosses that correspond to NOAA ARs are found at most up to -60$\degr$. Therefore, we conclude that the secondary peak observed in the distribution of $G_{S}$ is mostly due to newly emerging active regions that contain small umbrae contributing to this secondary peak. 

The Ising energy distribution (Figure~\ref{fig:histos}b) exhibits two clear peaks, one at the low end, at $log E_{Ising}=-0.6$ and another one at $log E_{Ising}=3.9$. In fact, it appears that the distribution consists of two components, an asymmetrical one at the higher end and a narrow, nearly symmetrical one, at the low end of the range. The latter is almost exclusively associated with SHARPs that do not correspond to NOAA ARs and are found $\pm$50$\degr$ from the central meridian. The low values produced for these regions are due to the fact that these regions are either plage/enhanced network regions or decaying active regions. In such regions, opposite polarities are separated by long distances and therefore, the corresponding $E_{Ising}$ attains low values.

Using magnetic partitions to calculate the Ising energy, alleviates the effect of non-active-regions to the distribution (Figure~\ref{fig:histos}c). The distribution is slightly asymmetric and shows a clear peak at $log E_{Ising,part}-2.8$. The NOAA AR SHARP-associated distribution (reen bars) is slightly asymmetric, as well, and the peak is shifted towards higher values, at $log E_{Ising,part}-2.2$, showing that well formed active regions that are visible in the continuum exhibit statistically higher $E_{Ising,part}$ values.

The distribution of the Ising energy of the umbrae, $E_{Ising,spot}$, is shown in Figure~\ref{fig:histos}d. It is similar to the distribution of $G_{S}$, only with a lower secondary peak at the lower range. It is asymmetric, with a dominant peak at $log E_{Ising,spot}=-2$. The corresponding distribution of the NOAA AR-associated HARPs peaks at the same values and is very similar to the total distribution. Since HARPS that contain no umbrae are eliminated, the two distributions are essentially referring to the same ``kind'' of regions of interest, only differing in the sample size (since NOAA-AR-assigned HARPs are fewer). 

For all four parameters, the distributions of the values that correspond to flaring regions are asymmetric and peak at the high end of the value range. This is another indication that for this sample flaring regions are statistically associated with higher values of the parameters. 

 \begin{figure}
 \centerline{\includegraphics[width=1\textwidth]{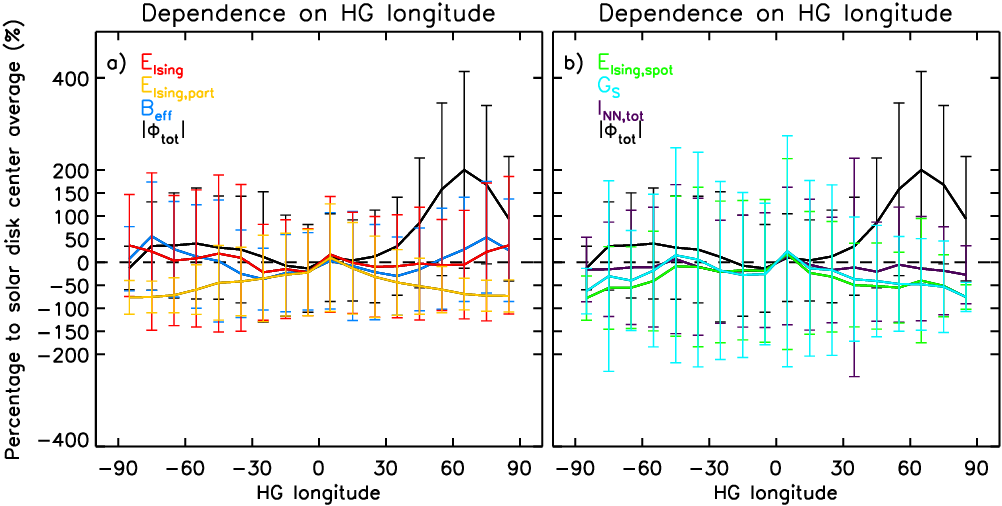}}
 \caption{Variation of the values of the seven morphological predictors (the four new ones, \textit{i.e.} $G_{S}$, $E_{Ising,}$, $E_{Ising,part}$, and $E_{Ising,spot}$, along with $B_{eff}$, $I_{NN,tot}$ and $\Phi_{tot}$), expressed as a percentage of the respective averages at the disk center. The average values are calculated in $10\degr$-wide bins and the error bars represent 1$\sigma$ from the average. The \textit{horizontal dashed line} marks zero percentage variation. Data are plotted in two panels to avoid overcrowding.} 
\label{fig:position}
 \end{figure}

The scatter-plots also contained in Figure~\ref{fig:histos} show the variation of the four parameters as a function of the heliographic longitude. No parameter shows dramatic decrease or increase towards the solar limb (\textit{i.e.} towards higher HG longitude). Furthermore, towards the limbs, all distributions appear slightly narrower. To better illustrate the HG dependence of the four parameters, we plot the average value of each predictors calculated in $10\degr$-wide bins as a percentage of the average value at the solar disk center (Figure~\ref{fig:position}). For comparison, we also add $B_{eff}$ (blue line), $I_{NN,tot}$ (purple line), and $\Phi_{tot}$ (black line). The latter shows the most dramatic variation with HG longitude. Towards the west limb, it may be up to 200\% higher than at the disk center. This asymmetry is attributed to the noise level of the detector \citep{hoeksema14}. 

$B_{eff}$ and $E_{Ising}$ (Figure~\ref{fig:position}a) also increase towards the limb but only up to 50\%. The similar behaviour of these two parameters was also shown in \citet{guerra18}. $E_{Ising,part}$ (Figure~\ref{fig:position}a), $E_{Ising,spot}$, and $G_{S}$ (Figure~\ref{fig:position}b), on the other hand, decrease by more the 50\% towards the limb. $I_{NN,tot}$ (Figure~\ref{fig:position}b) is the least affected parameter by solar disk position. It reduces, on average, by less than 50\% with increasing HG longitude. These four quantities also exhibit the least spread towards the limb, as demonstrated by the error bars in Figure~\ref{fig:position}.  

The four new parameters studied, as well as $B_{eff}$, are calculated here from the LOS component of the magnetic field (in line with their original definitions), which does not show the asymmetry of the longitudinal distribution of the $B_{r}$-derived $\Phi_{tot}$ values. A comparison made for $E_{Ising}$ and $B_{eff}$ in \citet{guerra18} showed that using the $B_{r}$ component results in higher values, on average, towards the limbs. This effect was more pronounced on $B_{eff}$, which depends not only on the distances between magnetic partitions but also on the values of the magnetic field, thus transferring the $B_{r}$ asymmetry to the values of $B_{eff}$ as well. Although, the vector magnetogram is utilized to calculate $I_{NN,tot}$, this predictor is relatively ``immune'' to the asymmetry of $B_{r}$. This is anticipated because only the $B_{\theta}$ and $B_{\phi}$ are used to calculate the vertical current density.

The increase of $B_{eff}$ and $E_{Ising}$ towards the limbs is due to the fake polarity inversion lines introduced in the LOS magnetograms by the geometric foreshortening of active regions towards the limb. The latter, however, also results in an overall reduced area and number of the magnetic partitions and detected umbrae and, thus, the values of $G_{s}$, $E_{Ising,part}$, and $E_{Ising,spot}$ reduce towards the limb.

 \begin{figure}
 \centerline{\includegraphics[width=1\textwidth]{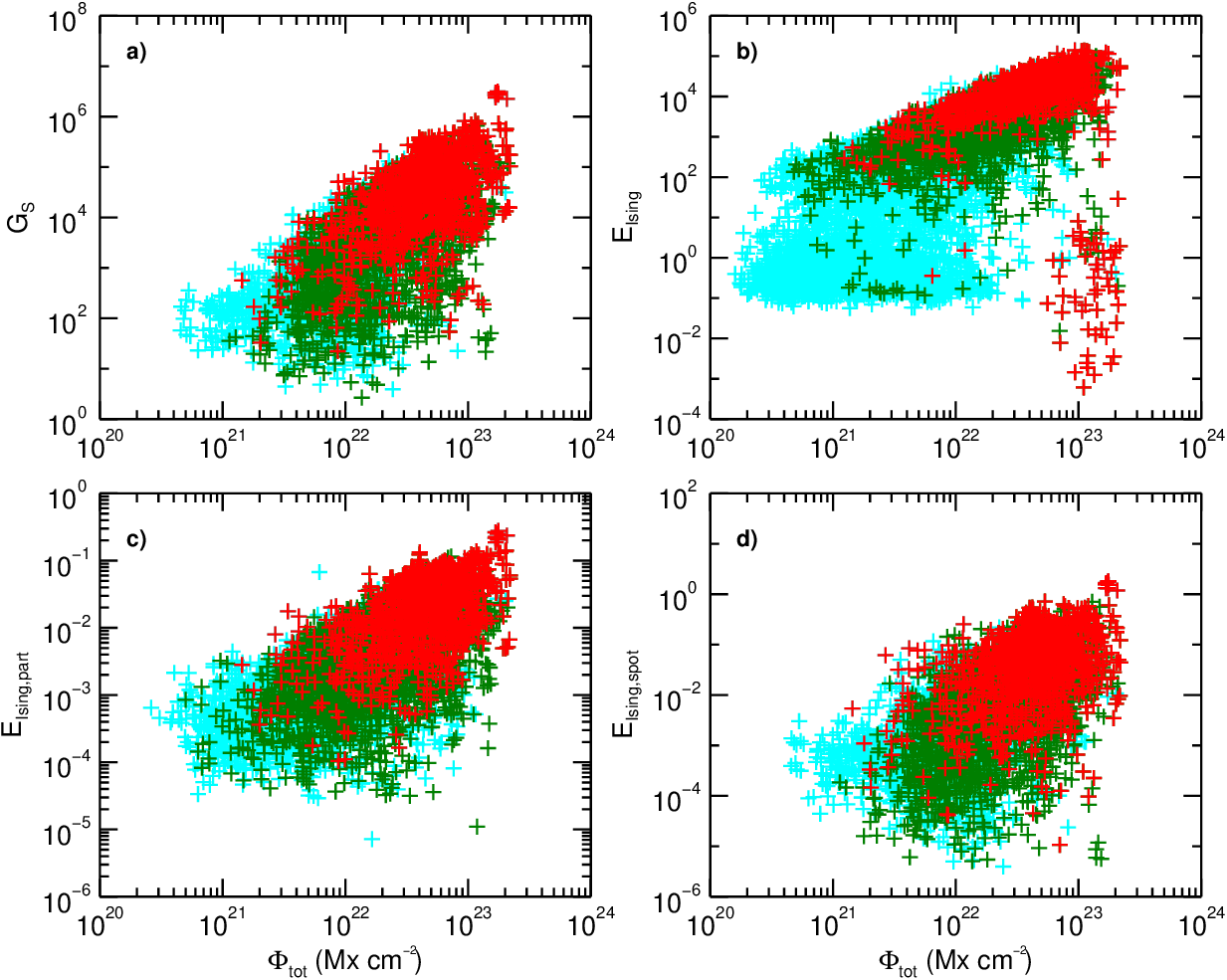}}
 \caption{Scatter plots between the four parameters and the total unsigned magnetic flux. \textit{green crosses} mark HARPs that contain NOAA AR while \textit{red crosses} mark flaring HARPs.}
\label{fig:params_flux}
 \end{figure}

\subsection{Parameter Correlations}
\label{s:params_corr}

The total unsigned magnetic flux, $\Phi_{tot}$, is a rudimentary  measure of the extent of an active region and in principle it also depends on the extent of the HARP cutout. In Figure~\ref{fig:params_flux} we plot the four parameters \textit{versus} the total unsigned magnetic flux. The corresponding correlation coefficients are found in Table~\ref{Table:t2}. For their calculation, only the non-zero values are taken into account (considering zero values as well, the correlation coefficients slightly increase without changing the essence of the results). The HARPs that correspond to NOAA AR (green crosses) have higher $\Phi_{tot}$ and thus the corresponding scatter plots are shifted towards the right. For $E_{Ising}$ (Figure~\ref{fig:params_flux}b), it is verified that the low-end of the distribution corresponds to regions with lower total unsigned magnetic flux, confirming that these are either plages or decaying active regions. 

In general, the four parameters are from significantly ($\gtrsim$0.5) to highly ($>>$0.5) correlated to $\Phi_{tot}$. The highest correlation with $\Phi_{tot}$ is found for $E_{Ising}$, ranging between 0.65 (if only NOAA AR HARPs are used) and 0.79 (if all HARPs are used). Apparently, $E_{Ising}$ is the parameter mostly affected by the extent of the region of interest, since a more extended region or FOV generally results in more opposite-polarity pixel pairs and, in principle, higher Ising energy. The parameters least correlated with $\Phi_{tot}$ are the ones that depend on the sunspot umbrae distribution, namely $G_{S}$ and $E_{Ising,spot}$.

 \begin{figure}
 \centerline{\includegraphics[width=1\textwidth]{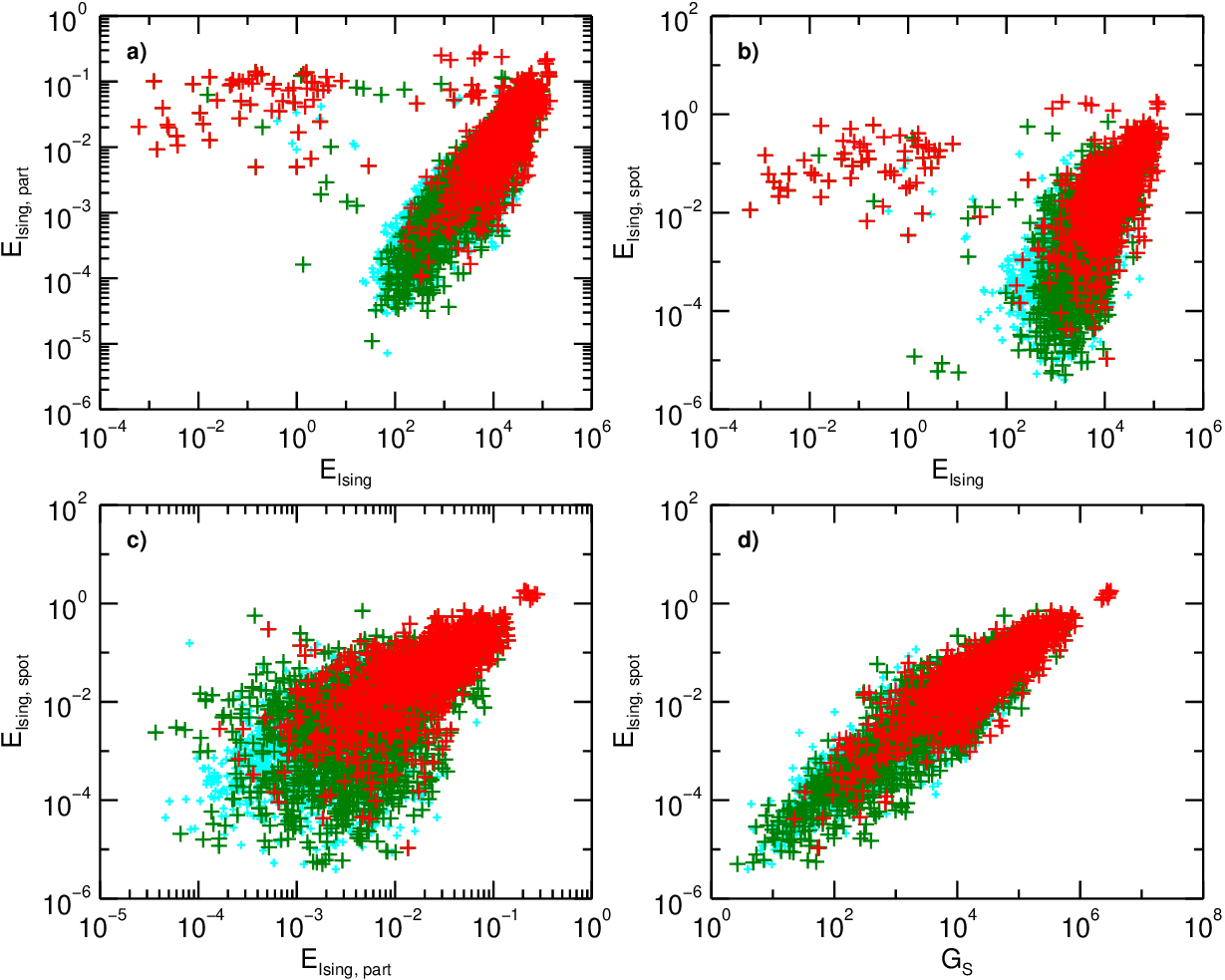}}
 \caption{Scatter plots between the four parameters. \textit{Green crosses} mark HARPs that contain NOAA AR while \textit{red crosses} mark flaring HARPs.}
\label{fig:ising_scat}
 \end{figure}

\begin{table}
\caption{Rank (linear) correlation coefficients between the four studied morphological parameters and the total unsigned magnetic flux ($\Phi_{tot}$). Also included are the respective values of the effective connected magnetic field strength ($B_{eff}$) and total non-neutralized currents ($I_{NN,tot}$).}
\setlength{\tabcolsep}{4.3pt}
\begin{tabular}{lccccc}
\hline
  &  &ALL HARPs\\
\hline
  &$G_{S}$&$E_{Ising}$&$E_{Ising,part}$&$E_{Ising,spot}$\\
\hline
$\Phi_{tot}$&0.72(0.40)&0.79(0.62)&0.78(0.61)&0.70(0.47)\\
$B_{eff}$&0.74(0.36)&0.81(0.76)&0.73(0.60)&0.67(0.43)\\
$I_{NN,tot}$&0.75(0.64)&0.79(0.70)&0.72(0.76)&0.72(0.70)\\
$E_{Ising,spot}$&0.87(0.90)&0.65(0.5)&0.62(0.76)& \\
$E_{Ising,part}$&0.63(0.70)&0.79(0.62)& & \\
$E_{Ising}$&0.75(0.39)& & & \\
\hline
  &  &NOAA AR\\
\hline
$\Phi_{tot}$&0.58(0.44)&0.65(0.49)&0.56(0.46)&0.54(0.37)\\
$B_{eff}$&0.74(0.58)&0.74(0.74)&0.63(0.62)&0.67(0.53)\\
$I_{NN,tot}$&0.76(0.80)&0.75(0.77)&0.73(0.76)&0.75(0.75)\\
$E_{Ising,spot}$&0.88(0.83)&0.67(0.66)&0.63(0.67)& \\
$E_{Ising,part}$&0.64(0.67)&0.74(0.65)& & \\
$E_{Ising}$&0.74(0.69)& & & \\
\hline
\end{tabular}
%\tablefoot{}
\label{Table:t2}
\end{table}

We also compare the four new parameters with $B_{eff}$ and $I_{NN,tot}$. The correlation coefficients are included in Table~\ref{Table:t2} while the corresponding scatter plots are not shown, for brevity. In general, all four parameters are highly correlated with $B_{eff}$ and $I_{NN,tot}$ and, again, $E_{Ising}$ exhibits slightly higher correlation. Restricting the analysis only to NOAA AR results in slightly higher correlation but does not change the results dramatically. By definition $I_{NN,tot}$ will produce zero results for active regions without strong MPILs while $B_{eff}$ is zero for regions that lack one polarity. This means that non-zero results represent largely the same populations, regardless of NOAA AR identification. In general, the correlation of the four new parameters with $B_{eff}$ and $I_{NN,tot}$ is higher than with $\Phi_{tot}$. This means that the four new parameters, in addition to reflecting the extent of an active region, also incorporate information related to MPILs. 

Figure~\ref{fig:ising_scat} contains scatter plots between the four parameters themselves. All parameters are positively correlated with relatively significant-to-high correlation coefficients. The correlation generally improves if we take into account only NOAA AR-associated HARPs. 

Out of the four parameters, $G_{S}$ and $E_{Ising,spot}$ are very highly correlated. Since these parameters both depend on the separation distance of opposite polarity umbrae, we believe that the inclusion of information concerning the magnetic field strength, through the fitting relation (Equation~\ref{equation:gs_calibr}), does not add extra information to the predictor. This result is in line with the analysis presented by \citet{korsos14} which shows that it is the decrease in the separation distance rather than the change in the flux that produces the increase in the horizontal magnetic gradient. This finding also means that the calibration relation probably does not affect drastically the efficiency of $G_{S}$.
\par

\subsection{Potential as Solar Flare Predictors}
\label{s:flare_predi}

In this section we will examine whether for this representative sample of HARP observations, the four new morphological predictors can offer an efficient way to separate flaring from non-flaring regions. Their efficiency will be judged in terms of comparison against the total unsigned magnetic flux, $\Phi_{tot}$, which is the simplest quantity one can deduce from active region observations and $I_{NN,tot}$ and $B_{eff}$.

We rely on Bayesian inference of flaring probabilities, following the approach by \citet{wheatland05} and \citet{geo12}. Consider a threshold value of a parameter. The Bayesian conditional probability, with condition contingent to the predictor threshold, for a SHARP magnetogram and the active region(s) it contains, is given by

\begin{equation}
{p=\frac{F+1}{N+2}},
\label{eq:bayes_prob}
\end{equation}

\noindent where F is the number of flaring regions and N the total number of regions with parameter value higher than the threshold. The corresponding uncertainty is given by:

\begin{equation}
{\delta p=\sqrt{\frac{p(1-p)}{N+3}}}.
\label{eq:bayes_error}
\end{equation}

\noindent Equation~\ref{eq:bayes_prob} simply provides the degree of segregation between flaring and non-flaring populations above a given threshold of the prediction parameter.

To facilitate comparisons between the seven parameters, we normalized each parameter with its maximum value and chose bins of values such that each bin had the same statistical weight. Note that for this analysis all values of the predictors are taken into account, both zero and non-zero, since they all bear a physical meaning.

 \begin{figure}
 \centerline{\includegraphics[width=1\textwidth]{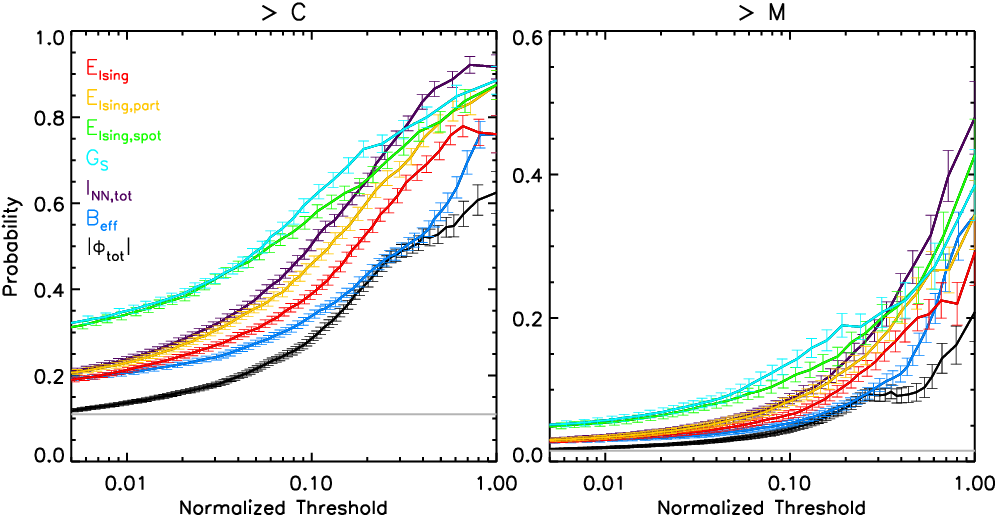}}
 \caption{Bayesian inferred flaring probabilities for flare class C and greater (\textit{left}) and flare class M and greater (\textit{right}). Horizontal \textit{gray lines} indicate the climatological frequency of the sample, \textit{i.e.} the overall ratio between the corresponding flaring and non-flaring populations.}
\label{fig:bayes}
 \end{figure}

In Figure~\ref{fig:bayes} we plot the Bayesian probabilities as a function of the normalized thresholds that correspond to each bin. All four parameters produce higher probabilities than the total unsigned magnetic flux, which means that they could be worth considering in an automated flare-prediction scheme, since they appear to achieve a better segregation between flaring and non-flaring HARP populations. Out of the four new predictors investigated here, the sum of the horizontal gradient of the magnetic field, $G_{S}$, along with the Ising energy of the sunspot umbrae, $E_{Ising,spot}$, produce almost the same probabilities and are the most efficient, with $G_{S}$ being slightly better. For larger than C class flares, these probabilities reach up to 0.85. For even larger flares, however, maximum probabilities remain low (up to $\sim$0.4 for M class events).

The efficiency of all four parameters is comparable to the efficiency of $I_{NN,tot}$ and $B_{eff}$. $I_{NN,tot}$ expresses the net currents that are injected into the corona and quantifies the presence of strong MPILs and, for this sample of observations, it produces the highest Bayesian probabilities. It should be noted that for the calculation of $I_{NN,tot}$ the vector of the magnetic field is required so this predictor encompasses more information. The efficiency of $B_{eff}$ (which requires only the LOS or the normal component of the magnetic field) in this sample is lower, although higher than the total unsigned magnetic flux and, in some cases, the original expression of $E_{Ising}$. 

In this tentative comparison between predictors (a comprehensive validation-based comparison will be achieved in the framework of FLARECAST) one may note the following two elements. First, that each predictor quantifies the presence of strong MPILs in a different manner. Using the partitions and the sunspot umbrae as interacting magnetic elements to calculate the Ising energy leads to improved probabilities compared with the original formulation of $E_{Ising}$. For $E_{Ising,spot}$ these probabilities are as high as the ones calculated for $G_{S}$. This result shows that, out of the four parameters, $E_{Ising,spot}$ and $G_{S}$ incorporate in the most efficient way the compactness of magnetic polarities at close proximity. This feature is associated with the presence of the PILs, convergence motions, and shear in flare-productive active regions. On the other hand, $B_{eff}$ quantifies PILs in a different way, dictated by a chosen magnetic connectivity, while the total unsigned non-neutralized currents represent the physical cause of PIL formation and shearing motions \citep{georgoulis12b}. 
Based on the findings of this section, we may conjecture that the efficiency of a predictor is as high as robust is its association with the physical cause of the flaring phenomenon, \textit{i.e.} the strong shearing motions along magnetic polarity inversion lines.

A second aspect that should be taken into account is the dependence of the predictor values on the solar disk position. Given that each predictor utilizes slightly different information, this effect is nontrivial. Comparing the results of this section with the findings in Section~\ref{s:params_distr}, it appears that the predictor which produces the highest flaring probabilities, that is $I_{NN,tot}$, is also the one least affected by projection effects. Furthermore, the flaring probabilities derived from those predictors with lower values and narrower distributions towards the limb ($E_{Ising,part}$, $E_{Ising,part}$, and $G_{S}$) are systematically higher. This is reasonable since, otherwise, increased predictor values and distribution spread towards the limb would result in an increase of the total number of regions above higher thresholds, $N$, thus resulting to decreased Bayesian probabilities for these thresholds. This is the case of $\Phi_{tot}$ and $B_{eff}$. 

In conclusion, regarding flare-prediction across the entire solar disk one may conjecture that an efficient predictor should combine physical intuition and immunity to projection effects. Ideally, a comparative analysis based on using the entire available sample of solar active region observations and predictors can be used to assess which of these predictors is more efficient. These should feed advanced (namely machine learning) prediction algorithms, with which one can then rank the importance of the predictors based on the algorithm performance-scores \citep[as \textit{e.g.}, in ][]{florios18}. This is, of course, outside the scope of the present study but is a core task of the FLARECAST project. 

\section{Discussion and Conclusions}
\label{s:conclusions}

Two proposed flare predictors, the sum of the horizontal magnetic gradient, $G_{S}$ \citep{korsos16} and the Ising energy \citep{ahmed10}, along with two new modifications of the Ising energy, were tested for the first time on a representative sample of Solar Cycle 24 SDO/HMI HARPs.

The temporal variation of the four parameters calculated for two different (\textit{i.e.} one intensely flaring and one non-flaring) active regions shows that they appear to reflect the development of morphological characteristics linked with increased flaring activity (PIL formation and emergence of magnetic flux). Overall, they exhibited significantly higher values for the flaring active region, especially prior to repetitive flaring activity.

From the sample of 9454 point-in-time observations, it was inferred that the four parameters show only a slight dependence on longitudinal position, which we attribute to their weak dependence on the magnetic field measurements themselves. Indeed, these parameters either decrease ($G_{S}$, $E_{Ising,part}$, $E_{Ising,spot}$) or marginally increase ($E_{Ising}$). It was also found that they are positively correlated with the size of the regions of interest, as expressed by the total unsigned magnetic flux, $\Phi_{tot}$. The original form of Ising energy exhibits the highest correlation, while the $G_{S}$ and $E_{Ising,spot}$ the lowest. Except $E_{Ising}$, the rest of the parameters produce zero values for plages of decaying active regions that either contain no sizeable magnetic partitions or no visible sunspots. Most of these regions do not correspond to NOAA AR and are scarcely associated with flares. Their high correlation with $B_{eff}$ and $I_{NN,tot}$ shows that the four new predictors indeed incorporate information associated with strong PILs.

Although all four new predictors are significantly-to-highly correlated, the particularly high correlation between $G_S$ and $E_{Ising,spot}$ probably means that they parameterize the morphological complexity of a given magnetic configuration in similar ways. They both depend on the inverse distance and they both rely on the presence of sunspot umbrae. For $G_{S}$, the inclusion of the magnetic field information, via the calibration relation seems to add no extra information. This is also in line with the analysis of \citet{korsos14} and \citet{korsos15} which show that it is the decrease in the distance between interacting umbrae that causes the increase in the value of $G_{S}$ that precede flares. 

By comparison with the corresponding efficiency of the total unsigned magnetic flux, it was found that all four tested parameters could be useful as flare predictors. The efficiency of the Ising energy, in terms of Bayesian-inferred flaring probabilities, is significantly improved when one considers the partitions of the magnetic flux or the sunspot umbrae as the interacting magnetic elements, instead of individual strong field pixels. Such a consideration is closer to the physical reality of interacting bundles of twisted flux tubes in the corona. This result is also corroborated through comparison with the corresponding efficiency of $I_{NN,tot}$ \citep{kontogiannis17}, a predictor that quantifies the net currents injected to the corona and is exclusively linked to the presence of sheared PILs \citep{georgoulis12b}. In conclusion, $E_{Ising,spot}$ and $G_{S}$ seem to be the most efficient potential predictors of the ones tested. 

Having demonstrated the potential efficiency of $E_{Ising,spot}$ we find it interesting that incorporating data from different databases, such as the Debrecen data, may also be useful in flare prediction. Perhaps in the future it might be useful to examine whether $E_{Ising,spot}$, $G_{S}$ or similar quantities can be calculated from synoptic ground-based observations and to what extent could they be utilized for a reliable flare forecasting. This would provide a very useful alternative in case the regular flow of high-quality magnetograms is temporarily disrupted or altogether interrupted, for any reason. 

Finally, we performed a preliminary comparison between the four new parameters and previously proposed, promising predictors. This analysis showed that two factors affect the efficiency of predictors: 1) how efficiently they quantify the presence of strong MPILs and 2) the different impact of projection effects and solar disk curvature on their calculation. As shown in Section~\ref{s:flare_predi}, the latter is not trivial but varies depending on the details of each calculation method. In this respect, $I_{NN,tot}$ is both insensitive to projection effects and robust in terms of physical intuition while $G_{S}$ and $E_{Ising,spot}$ appear to adequately quantify morphological information while largely overcoming projection effects. On the other hand, $B_{eff}$, a very promising and well-established predictor is, apparently, strongly affected by projection effects, which justifies its use in restricted heliographic longitudes.

As already noted, the new morphological predictors examined here were calculated for the LOS component of the magnetic field. This was chosen so because both $E_{Ising}$ and $G_{S}$ were introduced for LOS magnetograms. Different dependence on the position on the solar disk and consequently different flaring rates are expected in case the radial component of the magnetic field, $B_{r}$, is used.
This was illustrated for $B_{eff}$ and $E_{Ising}$ in \citet{guerra18}, where it is suggested that $B_{r}$ may be preferable in some cases. However, performing another comparison between $B_{r}$- and $B_{LOS}$-derived parameters exceeded the scope of this study. A more comprehensive assessment of the importance of all predictors so far proposed in the literature, calculated for both $B_{r}$ and $B_{LOS}$ (if applicable), may be achieved when such an extensive set of predictors is used with advanced statistical methods \citep[\textit{e.g.} machine learning, see][]{florios18}. This is one of the main tasks of the FLARECAST project that will provide definitive, fully validated, results in the near future.

 \begin{acknowledgements}
\noindent This research has been funded by the European Union's Horizon2020 research and innovation programme Flare Likelihood and Region Eruption Forecasting" (FLARECAST) project, under grant agreement No.640216. The data used are courtesy of NASA/SDO, the HMI science team, and the \textit{Geostationary Satellite System} (GOES) team. This work also used data provided by the MEDOC data and operations centre (CNES/CNRS/Univ. Paris-Sud), \url{http://medoc.ias.u-psud.fr/}. We thank the anonymous referees for their valuable comments.
\end{acknowledgements}

\textbf{Disclosure of Potential Conflict of Interest.} The authors declare that they have no conflict of interest.

\bibliographystyle{spr-mp-sola}
\bibliography{references}

\end{article}
\end{document}